RESEARCH ARTICLE

# Performance Evaluation of the WeatherEx Forecasting System (WFS) for Extreme Monsoon Rainfall over Kerala, 31 July-2 August 2026

**Rashad P E[1], Akshay Sunil[1]**

[1] *Experiqs Private Limited, India*

**Author contacts: Rashad P E, ribnummer@gmail.com; Akshay Sunil, akshay.sunil@experiqs.tech**

## Abstract

An extreme monsoon rainfall episode affected Kerala during 31 July-2 August 2026, with the 24-h station-interpolated analysis ending at 08:30 IST on 2 August showing very heavy to extremely heavy rainfall over northern Kerala and local maxima exceeding 350 mm. This study evaluates the archived operational performance of the WeatherEx Forecasting System (WFS), a regionally calibrated numerical forecasting framework coupled with AI/ML-based post-processing. Verification was performed using 108 common rain-gauge station pairs and spatial comparison with station-interpolated rainfall analyses. The operational 6 km WFS configuration produced an RMSE of 72.7 mm, an MAE of 55.1 mm and a mean bias of -24.1 mm. It maintained near-perfect probability of detection (POD) through the lower and moderate thresholds, retained POD = 0.83 at 100 mm, and exhibited negligible false-alarm measures through 100 mm. The 4 km diagnostic configuration had substantially larger bulk errors (RMSE = 149.6 mm; MAE = 119.9 mm), but a smaller mean dry bias (-9.7 mm) and greater sensitivity to localized extremes, with POD = 0.24 at 200 mm compared with 0.04 for the 6 km configuration. Spatial comparison indicates that WFS reproduced the Ghats-aligned heavy-rainfall corridor and sharper mesoscale intensity gradients than the archived global reference, although localized displacement remained a major error source. The results support a scale-aware operational strategy in which the 6 km product provides the more stable Kerala-wide baseline while finer-resolution guidance is used as an extreme-rainfall diagnostic layer. Because this analysis covers one high-impact event, the findings should be interpreted as event-level evidence rather than climatological skill.

## 1. Introduction

Extreme rainfall forecasting over Kerala is challenging because synoptic monsoon circulation, Arabian Sea moisture transport, coastal convergence, embedded convection and steep Western Ghats orography interact over short spatial and temporal scales. The operational difficulty is therefore not limited to predicting whether the state will experience an active monsoon spell; the more demanding problem is resolving the location and magnitude of narrow rainfall maxima that control flood, landslide and infrastructure impacts. The August 2018 Kerala flood episode demonstrated the consequences of errors in the spatial distribution and intensity of rainfall, and subsequent evaluations have shown that even established global prediction systems exhibit threshold-dependent limitations over the state [1,2].

The 31 July-2 August 2026 episode provides a useful operational test because the observed field contained a strong north-to-central Kerala rainfall gradient and localized extreme maxima.

In the station-interpolated analysis used here, Cherupuzha recorded 439 mm, Tatamala Estate in Wayanad 380 mm, and Ayyankunnu in Kannur 356.5 mm over the 24-h period ending 08:30 IST on 2 August. Such localized totals are difficult for models to reproduce exactly, particularly when verification is performed at individual gauge locations.

The WeatherEx Forecasting System (WFS) is designed as a regional forecasting framework in which high-resolution numerical guidance is combined with region-specific calibration and AI/ML-based post-processing. This report evaluates the end-to-end operational product rather than individual internal physical parameterization choices. The objectives are to (i) assess the spatial representation of the Kerala rainfall event; (ii) quantify continuous station-based errors; (iii) evaluate event detection across increasingly severe rainfall thresholds; (iv) compare the 6 km operational product with a 4 km diagnostic configuration; and (v) place the event-level performance in the context of published rainfall-verification studies over Kerala and the Indian monsoon region.

## 2. WFS forecasting framework and case-study data

### 2.1 Regional forecasting framework

India spans highly heterogeneous hydroclimatic regimes, including the Himalayan arc, the Indo-Gangetic Plain, the arid and semi-arid northwest, central monsoon regions, long tropical coastlines, the Western and Eastern Ghats, and the northeast. The atmospheric processes that control high-impact rainfall therefore vary substantially by region. Global models remain essential for large-scale boundary and synoptic information, but kilometre-scale regional guidance and local calibration can provide additional value where terrain, convection and coastal processes dominate the forecast error structure.

**Table 1. Functional components of the WFS framework evaluated in this report.**

| Component | Primary function | Verification relevance |
|---|---|---|
| Regional numerical core | High-resolution simulation of circulation, moisture transport, terrain forcing and precipitation. | Provides physically consistent regional forecast fields. |
| Kerala calibration | Region-specific tuning and observation-based evaluation. | Targets systematic local, coastal and terrain-related errors. |
| AI/ML post-processing | Data-driven correction and refinement of operational guidance. | Supports bias correction and local forecast refinement. |
| Verification layer | Comparison with collocated station observations and spatial rainfall analyses. | Quantifies continuous error, event detection and forecast limitations. |

### 2.2 Event and verification data

The case study covers the extreme monsoon spell affecting Kerala from 31 July to 2 August 2026, with emphasis on the 24-h accumulation ending at 08:30 IST on 2 August. The observational spatial reference is a station-interpolated rainfall analysis. Point verification uses 108 common station pairs available across the archived WFS comparisons. A deterministic global forecast is retained only as an external spatial benchmark; it is not used as observational truth and is not included in the station-based error statistics reported for WFS.

## 3. Verification methodology

Continuous and categorical metrics are used together because no single score adequately characterizes an extreme-rainfall forecast. RMSE emphasizes large point-wise errors; MAE describes the typical absolute error; and mean bias indicates the systematic wet or dry tendency. Categorical verification is constructed from a contingency table containing hits (H), misses (M), false alarms (FA) and correct negatives (CN). Probability of Detection (POD) measures the fraction of observed events correctly forecast, False Alarm Ratio (FAR) measures the fraction of forecast events that did not occur, and Probability of False Detection (POFD) measures false alarms relative to all observed non-events.

$$RMSE = \left[\frac{1}{N} \sum_{i=1}^{N} (F_i - O_i)^2\right]^{1/2}$$

$$MAE = \frac{1}{N} \sum_{i=1}^{N} |F_i - O_i|$$

$$\text{Mean Bias} = \frac{1}{N} \sum_{i=1}^{N} (F_i - O_i)$$

$$POD = H / (H + M)$$

$$FAR = FA / (H + FA)$$

$$POFD = FA / (FA + CN)$$

Here $F_i$ and $O_i$ denote forecast and observed rainfall at station i, and N is the number of common station pairs. Thresholds from 1 to 200 mm are examined to distinguish performance for ordinary rainfall, heavy rainfall and the rarest event maxima. For kilometre-scale precipitation, point verification can impose a substantial displacement penalty: an intense forecast core shifted by only a few grid cells may receive a large station-wise error even when its magnitude and mesoscale structure are realistic. For this reason, spatial-neighbourhood verification is recommended for future multi-event evaluation [4].

## 4. Results

### 4.1 Spatial representation of the rainfall field

Figure 1 compares the station-interpolated observation with two archived 6 km WFS lead products and the external global reference. The observed rainfall field is dominated by a pronounced northern Kerala maximum extending across Kannur and Wayanad, with a broader very heavy rainfall corridor along the windward side of the Western Ghats. Both WFS panels reproduce the existence and orientation of this high-rainfall corridor and generate sharper local gradients than the global field. The global reference captures the broader wet pattern but is noticeably smoother and suppresses the most localized intensity maxima.

The spatial comparison also illustrates why evaluation must separate pattern skill from exact point agreement. WFS places strong rainfall in the correct broad physiographic corridor, but the location and width of individual high-intensity cells differ from the interpolated observation. Such displacement contributes directly to the station-wise RMSE and becomes increasingly important as model resolution increases.

### 4.2 Station-based continuous verification

Across the 108 common station pairs, the operational 6 km configuration produced RMSE = 72.7 mm and MAE = 55.1 mm, compared with 149.6 mm and 119.9 mm for the 4 km diagnostic product (Table 2). Relative to the 4 km run, the 6 km configuration reduced RMSE by approximately 51% and MAE by approximately 54%, indicating substantially more stable station-to-station agreement for this event. The 6 km mean bias was -24.1 mm, however, compared with -9.7 mm at 4 km, showing that the coarser operational field had a stronger systematic dry tendency.

The smaller mean bias of the 4 km product should not be interpreted as superior overall accuracy. Bias can be reduced through cancellation between local overprediction and underprediction, whereas the much larger RMSE and MAE at 4 km indicate greater spatial error variance. The results therefore point to a resolution-dependent trade-off: the 6 km field is the stronger bulk predictor, whereas the 4 km field retains more localized intensity but is more sensitive to displacement error.

**Table 2. Continuous station-based verification for 108 common station pairs.**

| WFS product | RMSE (mm) | MAE (mm) | Mean bias (mm) |
|---|---|---|---|
| 6 km operational | 72.7 | 55.1 | -24.1 |
| 4 km high-resolution diagnostic | 149.6 | 119.9 | -9.7 |

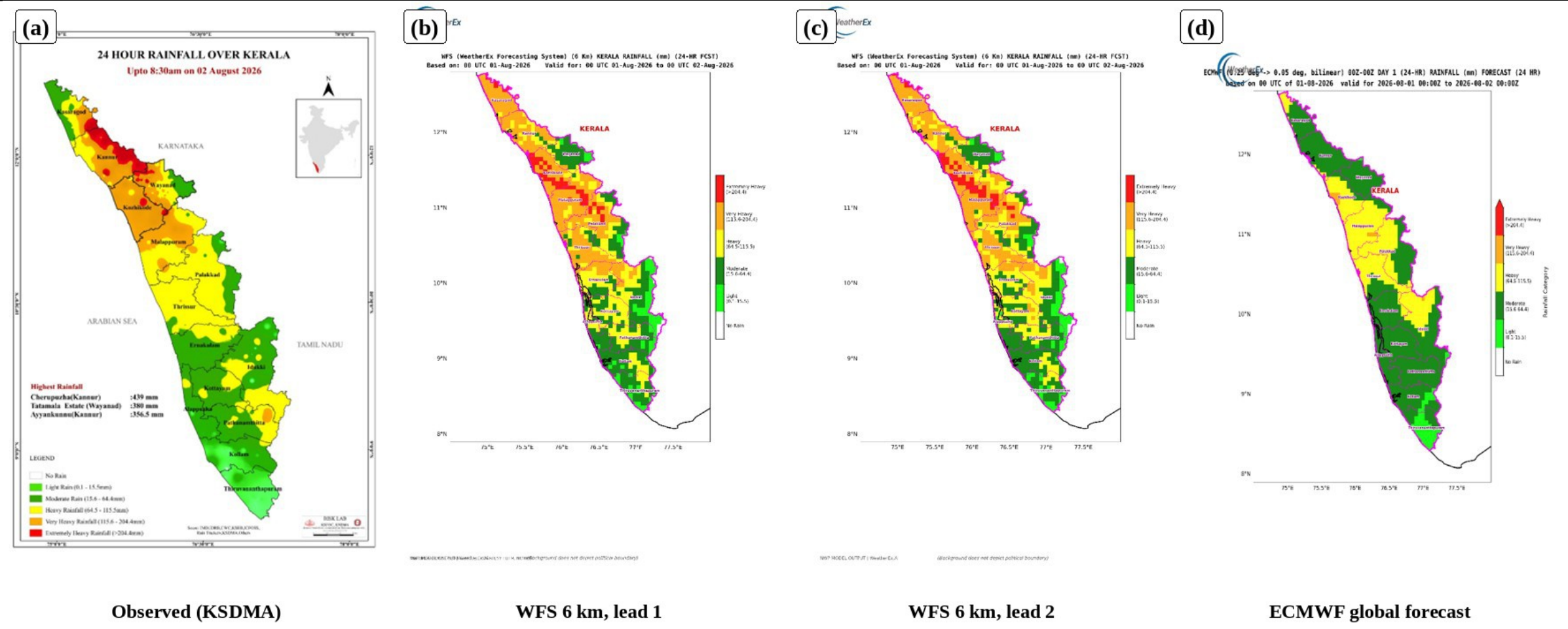

Figure 1. Spatial comparison of observed and forecast 24-h rainfall over Kerala for the period ending 08:30 IST on 2 August 2026. (a) KSDMA station-interpolated rainfall analysis, with the strongest observed rainfall over northern Kerala and local maxima of 439 mm at Cherupuzha, 380 mm at Tatamala Estate (Wayanad), and 356.5 mm at Ayyankunnu (Kannur). (b-c) Archived 6 km WFS lead-1 and lead-2 guidance, which reproduce the north-to-central Ghats-aligned heavy-rain corridor and retain localized very heavy to extremely heavy rainfall cores. (d) ECMWF global guidance for the same valid period, which captures the broad wet spell but presents a smoother field with weaker localized intensity. Observation and model fields have different native resolutions and interpolation characteristics; the figure is therefore interpreted primarily as a spatial-pattern comparison rather than a pixel-wise equivalence test.

### 4.3 Threshold-based detection: 6 km operational product

The 6 km product shows strong event detection through the heavy-rain regime (Figure 2). POD remains close to unity at the lower thresholds, is approximately 0.96 at 50 mm and 0.83 at 100 mm, while FAR and POFD remain effectively zero through 100 mm. This combination is operationally important because it indicates that, for this event sample, the system detected the great majority of stations exceeding moderate to heavy rainfall thresholds without generating a comparable number of false alarms.

Skill degrades at the highest thresholds. POD falls to about 0.58 at 150 mm and 0.04 at 200 mm, while FAR rises sharply. The decline is expected because the number of observed exceedances becomes small and the rainfall maxima are spatially compact. Thus, the 6 km product provides a strong baseline for broad heavy-rain detection but is not sufficient by itself for exact prediction of the rarest localized extremes.

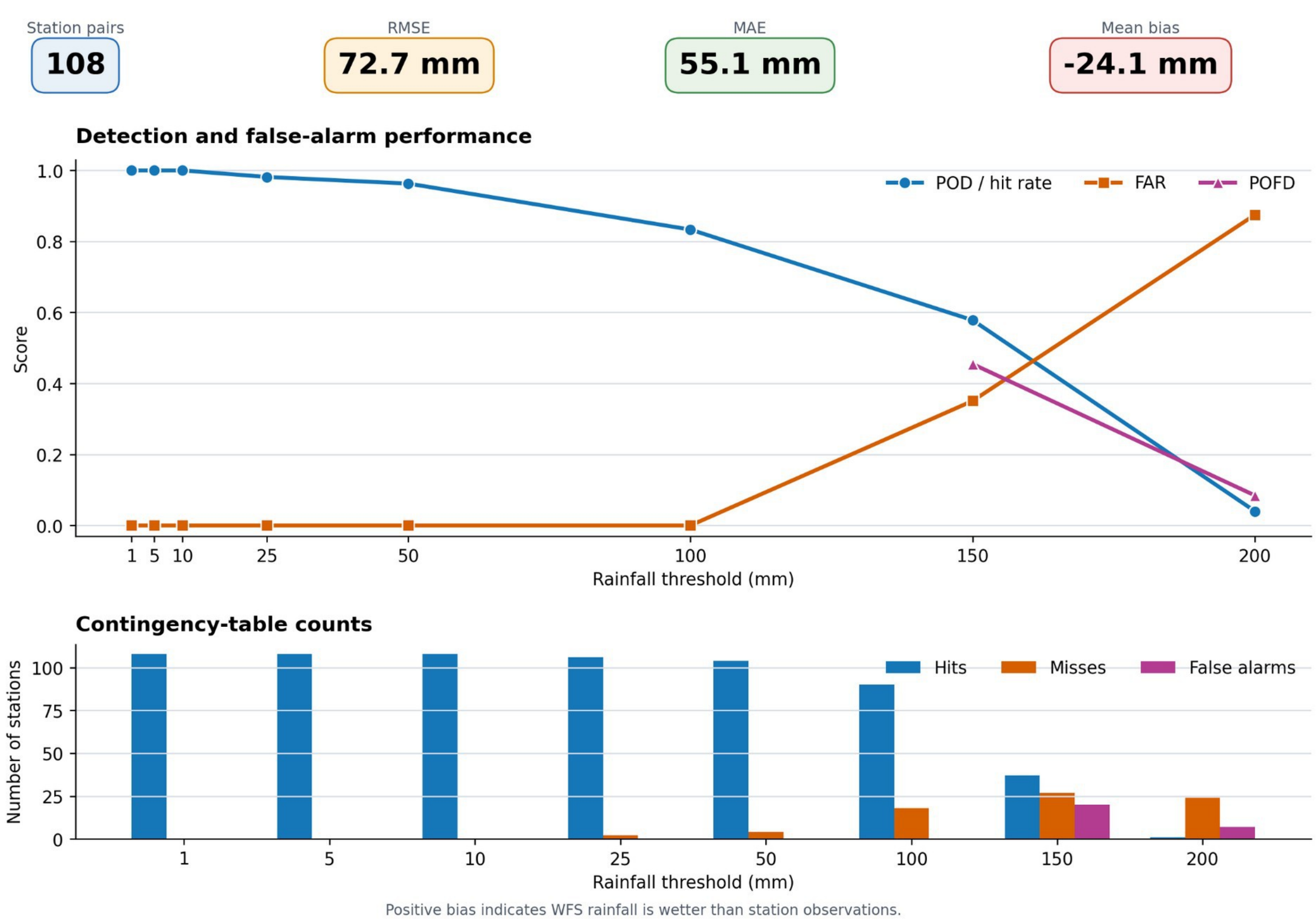

Figure 2. Threshold-based categorical verification of the operational 6 km WFS product against 108 collocated station observations. The upper panel shows POD, FAR and POFD as the rainfall threshold increases from 1 to 200 mm; the lower panel shows the corresponding contingency-table counts. POD remains near 1.0 through the lower thresholds, is approximately 0.96 at 50 mm and 0.83 at 100 mm, then decreases rapidly at 150-200 mm. FAR remains negligible through 100 mm and increases only at the rarest thresholds. The figure therefore demonstrates strong broad-event and heavy-rain detection while also identifying the principal limitation of the 6 km product: localized extreme rainfall above about 150-200 mm is frequently missed or displaced.

### 4.4 Threshold-based detection: 4 km diagnostic product

The 4 km diagnostic configuration has a different error structure (Figure 3). Its POD decreases more gradually with threshold, from approximately 0.91 at 25 mm to 0.73 at 50 mm and 0.56 at 100 mm. Although this is lower than the 6 km operational product at moderate and heavy thresholds, the 4 km configuration retains substantially more signal at the extreme end: POD = 0.24 at 200 mm compared with 0.04 at 6 km. This behavior is consistent with a finer grid preserving localized rainfall peaks that are smoothed in a coarser field.

The benefit at the extreme threshold comes at the cost of larger bulk error and increasing false alarms at 150-200 mm. The 4 km field is therefore best interpreted as a diagnostic refinement layer rather than a replacement for the 6 km operational baseline. In practice, the two products convey complementary information: the 6 km field provides more stable regional coverage and the 4 km field provides additional sensitivity to localized high-intensity cores.

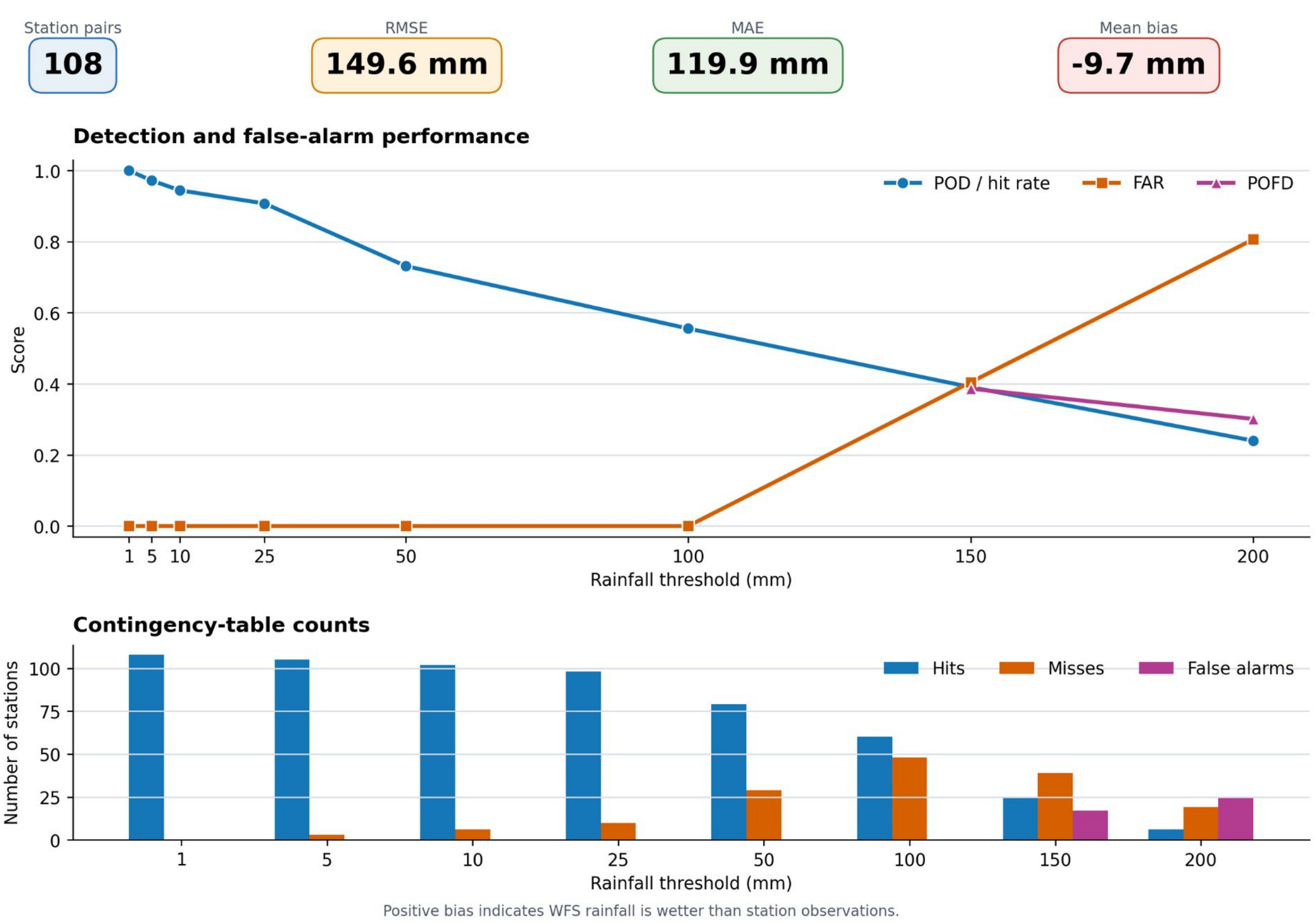

Figure 3. Threshold-based categorical verification of the 4 km high-resolution WFS diagnostic product against the same 108 station observations. POD decreases more steadily with threshold than in Figure 2, reaching about 0.73 at 50 mm, 0.56 at 100 mm, 0.39 at 150 mm and 0.24 at 200 mm. The 200 mm POD is substantially higher than for the 6 km product, indicating greater retention of localized extreme-rainfall signal. However, RMSE and MAE are much larger and FAR rises at the highest thresholds, showing that the added spatial detail is accompanied by stronger displacement sensitivity and more variable station-scale error.

## 5. Discussion

### 5.1 Resolution-dependent skill and displacement error

The results demonstrate why extreme-rainfall forecast quality cannot be summarized by a single metric. The 6 km product is clearly superior in RMSE and MAE and in categorical detection through 100 mm, but it has a stronger dry bias and very limited POD at 200 mm. The 4 km product reduces the mean dry bias and improves detection of the most localized extremes, yet its larger RMSE and MAE indicate that increased spatial detail does not automatically translate into improved point-wise accuracy.

This is a well-known feature of high-resolution precipitation verification. Finer grids can produce more realistic and intense rainfall structures while incurring a double penalty when a localized forecast feature is displaced: the model is penalized both where rain was forecast but did not occur and where rain occurred but was not forecast. Scale-selective verification methods such as the Fractions Skill Score were developed specifically to diagnose this problem by evaluating whether a forecast is useful over neighbourhood scales rather than requiring exact grid-point coincidence [4]. Future WFS validation should therefore combine station metrics with FSS or related neighbourhood exceedance scores, particularly at 100-200 mm thresholds.

### 5.2 Comparison with published Kerala and Indian rainfall verification

Nitha et al. [1] evaluated six operational global models over Kerala for the 2018-2022 southwest monsoon seasons. JMA produced the lowest Day-1 domain-mean RMSE and the highest anomaly correlation in that multi-season study, while ECMWF showed the most consistently balanced categorical performance, combining high POD with a moderate FAR of about 0.41 and the highest equitable threat score among the evaluated systems. Mandal et al. [3] similarly showed that categorical rainfall skill over the Indian monsoon region degrades as thresholds increase, especially in wetter regimes.

The present WFS results are encouraging in that the 6 km operational system maintained very high detection through 50 mm and POD = 0.83 at 100 mm with negligible false alarms through the lower thresholds. However, a strict numerical ranking against Nitha et al. is not methodologically valid: their statistics were based on multi-season verification on a 0.25-degree grid, whereas the present analysis is a single extreme event using 108 point-station pairs and different thresholds. The correct interpretation is therefore that WFS demonstrates event-level added value and a competitive heavy-rain detection signal, not that it has been proven universally superior to every global model.

**Table 3. Contextual comparison with published forecast-verification studies. Values are not directly rankable because the samples, grids and threshold definitions differ.**

| Study/system | Key result | Interpretation |
|---|---|---|
| Present WFS, 6 km | RMSE 72.7 mm; POD 0.83 at 100 mm | Lower bulk error and strong heavy-rain detection; dry bias and weak 200 mm detection. |
| Present WFS, 4 km | RMSE 149.6 mm; POD 0.24 at 200 mm | Better retention of localized extremes, but much larger displacement-sensitive error. |

| Study/system | Key result | Interpretation |
|---|---|---|
| Nitha et al. (2025) [1] | JMA Day-1 RMSE 14.73 mm; ECMWF high POD, FAR ~0.41, highest ETS | Useful multi-season Kerala benchmark, but not directly comparable with this event-level station analysis. |
| Mandal et al. (2007) [3] | Skill decreases as rainfall thresholds increase | Supports the threshold-dependent interpretation of WFS performance. |

### 5.3 Operational interpretation

For routine Kerala-wide guidance, the 6 km configuration is the stronger baseline because it combines markedly lower bulk error with high detection through the 100 mm threshold. The 4 km product should be interpreted as an extreme-event diagnostic layer that can preserve localized high-intensity peaks that are smoothed at 6 km. In flood and landslide applications, the most useful operational architecture is therefore multi-scale: regional confidence from the 6 km product, localized extreme-risk refinement from finer-resolution guidance, and observational/nowcasting updates as the event approaches.

The results also support a hybrid forecasting architecture, but the incremental contribution of AI/ML cannot be quantified from the present archive. A scientifically complete assessment should verify the raw numerical forecast and the post-processed product separately at identical stations, lead times and thresholds. That experiment would establish whether the AI/ML stage reduces bias and RMSE, improves reliability, or mainly redistributes spatial error.

## 6. Limitations and future validation

This analysis is event-based and should not be interpreted as a climatological estimate of WFS skill. The sample contains 108 common station pairs for one high-impact episode. Point gauges also have representativeness limitations when compared with kilometre-scale area-average model precipitation, especially over steep terrain. In addition, the available archive does not provide a controlled raw-versus-post-processed experiment, so the independent contribution of AI/ML cannot be isolated.

Future validation should extend across multiple monsoon seasons, lead times, rainfall regimes and physiographic zones. The recommended verification suite includes bias, RMSE, MAE, correlation, POD, FAR, CSI/ETS, reliability measures and neighbourhood-based spatial scores such as FSS. Lead-time-specific confidence intervals and event-stratified results are needed to determine whether the event-level skill reported here persists under routine operations. A second priority is independent verification using observation datasets not involved in calibration or post-processing.

## 7. Conclusions

The 31 July-2 August 2026 Kerala event shows that WFS provided a useful operational representation of a high-impact monsoon rainfall episode. The system captured the state-scale wet signal, the Ghats-aligned rainfall corridor and localized high-intensity structure that was substantially smoother in the archived global reference. The 6 km operational product provided the strongest overall station-based performance, with RMSE = 72.7 mm, MAE = 55.1 mm and POD = 0.83 at 100 mm. The 4 km diagnostic product had substantially larger bulk error but retained more of the localized extreme-rainfall signal, reaching POD = 0.24 at 200 mm compared with 0.04 at 6 km.

These results support a scale-aware regional forecasting strategy rather than a single-resolution interpretation. The 6 km field is the more robust baseline for Kerala-wide operational guidance, while finer-resolution products can add value for localized extreme-risk diagnosis. The next scientific step is to establish multi-event and lead-time-dependent skill, quantify the independent benefit of AI/ML post-processing, and adopt neighbourhood verification so that realistic but slightly displaced rainfall structures are not treated as complete forecast failures. The present study therefore provides evidence of operational added value for this event while retaining appropriate caution about generalization beyond the available sample.

**Data statement**

This report is based on the archived WeatherEx Kerala verification material for 31 July-2 August 2026. Numerical values and threshold statistics are retained from that archive. Internal physical-scheme names have been removed from the spatial figure so that the document focuses on end-to-end WFS performance rather than individual parameterization choices.

**Declaration of interest**

WFS is developed and operated by WeatherEx.ai, Experiqs Pvt. Ltd. The present document is an internal operational-performance evaluation. Independent multi-event verification is recommended before using these event-level results as a climatological performance claim.